\documentstyle[epsf,12pt]{article}
\def\bc{\begin{center}}
\def\ec{\end{center}}
\def\be{\begin{equation}}
\def\ee{\end{equation}}

\newcommand{\nn}{\nonumber}

\newcommand{\MSbar}{\overline{\mbox{MS}}}
\newcommand{\bea}{\begin{eqnarray}}
\newcommand{\eea}{\end{eqnarray}}

\begin{document}
\pagestyle{empty} 
\begin{flushright}
Edinburgh 98/8 \\
FTUV/98-40 and IFIC/98-41\\
ROME1-1209/98 \\
ROM2F/98/16 \\
SNS/PH/1998-011 \\
SWAT/190
\end{flushright}
\vskip 0.2 cm
\centerline{\LARGE{\bf{$B$-parameters for $\Delta S = 2$}}}
\centerline{\LARGE{\bf{Supersymmetric Operators}}}
\vskip 0.3cm
\centerline{\bf{C.R.~Allton$^a$, L. Conti$^b$, A. Donini$^c$, 
                V. Gimenez$^d$,  }} 
\centerline{\bf{L. Giusti$^e$, G. Martinelli$^f$, 
                M. Talevi$^g$, A. Vladikas$^b$}}
\vskip 0.3cm
\centerline{$^a$ Dep. of Physics, University of Wales Swansea, Singleton
Park,}
\centerline{Swansea, SA2 8PP, United Kingdom.}
\centerline{$^b$ INFN, Sezione di Roma II, and 
Dip. di Fisica, Univ. di Roma ``Tor Vergata'',}
\centerline{Via della Ricerca Scientifica 1, I-00133 Roma, Italy.}
\smallskip
\centerline{$^c$ Dep. de Fisica Teorica, Univ. Autonoma Madrid,}
\centerline{ Fac. de Ciencias, C-XI, Cantoblanco, E-28049 Madrid, Spain. }
\smallskip
\centerline{$^d$ Dep. de Fisica Teorica and IFIC, Univ. de Valencia,}
\centerline{Dr. Moliner 50, E-46100, Burjassot, Valencia, Spain.}
\smallskip
\centerline{$^e$ Scuola Normale Superiore, P.zza dei Cavalieri 7 and}
\centerline{INFN, Sezione di Pisa, 56100 Pisa, Italy.}
\smallskip
\centerline{$^f$ Dip. di Fisica, Univ. di Roma ``La Sapienza'' and
INFN, Sezione di Roma,}
\centerline{P.le A. Moro 2, I-00185 Roma, Italy.}
\smallskip
\centerline{$^g$  Department of Physics \& Astronomy, University of Edinburgh,}
\centerline{The King's Buildings, Edinburgh EH9 3JZ, UK.}
\vskip 1.0cm

\abstract{We present a calculation of  the matrix elements of the most general set
of $\Delta S=2$ dimension-six  four-fermion operators. The 
values of the matrix elements  are given  in terms of the corresponding 
$B$-parameters. Our results  can be used in many
phenomenological applications, since the operators considered here
 give important  contributions to $K^0$--$\bar K^0$ mixing  
in several  extensions of  the Standard Model (supersymmetry, 
left-right symmetric  models, multi-Higgs models etc.). 
The  determination of the matrix elements improves the accuracy
of the  phenomenological analyses intended to put bounds on basic parameters
of the different models, as for  example the pattern of the sfermion 
mass matrices. The calculation has been performed on the lattice,
using the tree-level improved Clover action at two different values
of the strong coupling constant ($\beta=6/g_0^2(a)=6.0$ and 
$6.2$, corresponding to  $a^{-1}= 2.1$ and 
$2.7$ GeV respectively),  in  the quenched approximation. 
The renormalization  constants and mixing coefficients of the lattice operators
have been obtained non-perturbatively.}
\vfill\eject
\pagestyle{empty}\clearpage
\setcounter{page}{1}
\pagestyle{plain}
\newpage 
\pagestyle{plain} \setcounter{page}{1}

\section{Introduction}
\label{sec:intro}
Important information on the physics beyond the Standard Model (SM),
such as   
supersymmetry,  left-right symmetric  models, multi-Higgs models etc., can be
obtained by studying FCNC processes. Among these, $\Delta F=2$ 
transitions play a very important role.  They have been 
used in ref.~\cite{susy,susybag},  for example,  to put constraints on the 
sfermion mass matrix.  In this paper we present the 
results of a lattice calculation of the matrix elements of the most 
general set of $\Delta S=2$ dimension-six  four-fermion operators, 
renormalized non-perturbatively in the RI (MOM) 
scheme~\cite{NP}--\cite{4ferm_teo}. 
Our results can be  combined with the recent two-loop  calculation of  the anomalous 
dimension matrix in the same renormalization scheme~\cite{scimemi}  to 
obtain $K^{0}$--$\bar K^{0}$ mixing amplitudes which are consistently 
computed at the next-to-leading  order.  A phenomenological application of the results 
for the matrix elements  given below, combined with a complete 
next-to-leading order  (NLO) 
evolution of the Wilson coefficients, will be presented 
elsewhere~\cite{inprep}. 
\par $K^0$--$\bar K^0$ mixing induces the neutral kaon mass difference 
$\Delta M_{K}$ and is related to the indirect CP violation parameter
 $\epsilon_K$. In the Standard Model, this transition occurs
via the dimension-six four-fermion operator $O^{\Delta S = 2}$, with
a ``left-left" chiral structure. 
The $B$-parameter of the matrix element $\langle \bar K^0 \vert O^{\Delta S = 2}
 \vert K^0 \rangle$,  commonly known as $B_K$, has been extensively studied on the lattice
due to its phenomenological relevance~\cite{sharpe:lat96}, and used in many 
phenomenological studies~\cite{parodi}. For the other 
operators, instead,   all the phenomenological analyses beyond the 
SM have used 
$B$-parameters equal to one, which in some cases, as will be shown 
below, is a very crude approximation.  
The present work is the first fully non-perturbative study  of the matrix 
elements of the complete set of $\Delta S=2$  four-fermion operators.  
With respect to other calculations, the systematic errors in our results are
reduced in two ways:
i) by using the   tree-level improved Clover action~\cite{sw} 
and operators~\cite{heat}  we obtain matrix elements for which 
discretization errors are of   ${\cal O}(\alpha_{s} a)$;  ii)  by   renormalizing 
non-perturbatively the lattice operators, we have  eliminated the 
systematic error due to the bad behaviour of lattice perturbation 
theory.  With the non-perturbative renormalization,
the  residual error is that  due to the truncation of the continuum perturbative 
series in the evaluation of the Wilson coefficients of the effective Hamiltonian. 
This error  
is  of ${\cal O}(\alpha_{s}^{2})$, where $\alpha_{s}$ is the 
continuum renormalization constant and could in principle be reduced 
by the calculation of the N$^{2}$LO corrections in the continuum. 
In  the RI scheme, for example,  this step has
recently been done for the calculation of the renormalized quark 
mass~\cite{fralu}. 
\par For the reader who is not interested in  technical details, 
 we now give the results for the $B$-parameters of the relevant 
operators.  The choice of the basis is arbitrary, and different 
bases can be found in the literature, see for example ~\cite{susy} 
and \cite{scimemi}. We have used the SUSY basis  which is also the one
for which the  numerical values of the Wilson  coefficients, computed at the NLO,  
will be given~\cite{inprep}, namely~\footnote{
We use here the Euclidean notation.}
\bea O_1 &=& \bar s^\alpha \gamma_\mu (1- \gamma_{5} ) d^\alpha \ 
\bar s^\beta \gamma_\mu (1- \gamma_{5} )  d^\beta ,  \nn \\ 
O_2 &=& \bar s^\alpha (1- \gamma_{5} ) d^\alpha \ 
 \bar s^\beta  (1- \gamma_{5} )  d^\beta ,  \nn \\ 
O_3&=& \bar s^\alpha  (1- \gamma_{5} )  d^\beta  \ 
 \bar s^\beta   (1- \gamma_{5} ) d^\alpha ,  \label{eq:ods2} \\ 
O_4 &=& \bar s^\alpha  (1- \gamma_{5} ) d^\alpha \  
\bar s^\beta  (1 + \gamma_{5} )  d^\beta ,  \nn \\ 
O_5&=& \bar s^\alpha  (1- \gamma_{5} )  d^\beta \ 
 \bar s^\beta (1 +  \gamma_{5} ) d^\alpha , \nonumber
 \eea 
where $\alpha$ and $\beta$ are colour indices. 
The $B$-parameters for these operators are defined as
\bea \langle  \bar K^{0} \vert \hat  O_{1} (\mu) \vert K^{0} 
\rangle &=&
\frac{8}{3} M_{K}^{2} f_{K}^{2} B_{1}(\mu) , \nn \\
\langle  \bar K^{0} \vert \hat O_{2} (\mu) \vert K^{0} \rangle &=&
-\frac{5}{3} \left( \frac{ M_{K} }{ m_{s}(\mu) + m_d(\mu) }\right)^{2}
M_{K}^{2} f_{K}^{2} B_{2}(\mu) , \nn \\
\langle  \bar K^{0} \vert \hat O_{3} (\mu) \vert K^{0} \rangle &=&
\frac{1}{3} \left( \frac{ M_{K} }{ m_{s}(\mu) + m_d(\mu) }\right)^{2}
M_{K}^{2} f_{K}^{2} B_{3}(\mu) , \label{eq:bpars}  \\
\langle  \bar K^{0} \vert \hat O_{4} (\mu) \vert K^{0} \rangle &=&
2 \left( \frac{ M_{K} }{ m_{s}(\mu) + m_d(\mu) }\right)^{2}
M_{K}^{2} f_{K}^{2} B_{4}(\mu) ,\nn \\
\langle  \bar K^{0} \vert \hat  O_{5} (\mu) \vert K^{0} \rangle &=&
\frac{2}{3} \left( \frac{ M_{K} }{ m_{s}(\mu) + m_d(\mu) }\right)^{2}
M_{K}^{2} f_{K}^{2} B_{5}(\mu) , \nn  \eea
where the notation $\hat O_{i}(\mu)$ (or simply $\hat O_{i}$) denotes the 
operators renormalized at the scale $\mu$.
A few words of explanation are necessary at this point. In 
eq.~(\ref{eq:bpars}) operators and  quark masses   are 
renormalized at the scale 
$\mu$  in the same scheme  (e.g. RI, $\MSbar$, etc.).  The numerical 
results for the $B$-parameters, $B_{i}(\mu)$  computed in this paper refer
 to the RI scheme. 
Moreover, without loss of generality,  we have omitted terms,
present in the usual definition of the $B$s,  which 
are of higher order in the chiral expansion. Since the definition of the 
$B$-parameters is conventional, we prefer 
to use those in eq.~(\ref{eq:bpars})  for which,
 as explained in sec.~\ref{sec:res},  the scaling properties
are the simplest ones.  
\par Our best estimates of the $B$-parameters, for a renormalization scale 
of $\mu =2$ GeV are 
\bea
B_{1}(\mu) &=& 0.69 \pm 0.21 , \nn \\
B_{2}(\mu) &=& 0.66 \pm 0.04 , \nn \\
B_{3}(\mu) &=& 1.05 \pm 0.12  , \label{eq:fres} \\
B_{4}(\mu) &=& 1.03 \pm 0.06 , \nn \\
B_{5}(\mu) &=& 0.73 \pm 0.10 \ . \nn \eea
\par 
The remainder of the paper is organized as follows:  in sec.~\ref{sec:npm}, we address the problem of operator
mixing  and renormalization and give a brief account of the 
non-perturbative method (NPM) used in   the
computation of the operator renormalization constants; in sec.~\ref{sec:res}, we
discuss the definition of 
the $B$-parameters and describe their extraction from the lattice correlation
functions; in sec.~\ref{sec:numres}, we 
present our results for the full operator basis, at $\beta=6.0$ and 
$6.2$ and for different  renormalization scales; a discussion of the 
errors assigned to the final results in eq.~(\ref{eq:fres}) can also be 
found in this section;  finally, in sec.~\ref{sec:concl}, we present  our conclusions.
\section{Non-Perturbative renormalization}
\label{sec:npm}
In this section, we briefly recall the reasons for which the non-perturbative 
renormalization of the lattice operators is important and describe the 
procedure which has been used to obtain, for the cases of interest,   
finite matrix  elements from the bare lattice 
operators. \par 
The Wilson lattice regularization breaks chiral symmetry. This implies
that each operator in the $\Delta S = 2$ Hamiltonian 
mixes with operators belonging to different chiral
representations~\cite{MARTIW,octet}. 
Because of the mixing induced by the lattice,
the correct chiral behaviour of the operators is
achieved with Wilson fermions only in the continuum limit. 
This represented a long-standing problem in the evaluation of 
$B_K$~\cite{BERNARD2,GAVELA}~\footnote{
In the staggered fermion approach, where chiral symmetry is
partially preserved, the $\Delta S = 2$ matrix element displays the
correct chiral behaviour. Thus, the $B_K$-parameter obtained with staggered
fermions \cite{SHARPE} has been deemed more reliable.}, only recently
solved with the introduction of Non-Perturbative Renormalization methods.
In these approaches the  renormalization constants (mixing matrix) are computed
non-perturbatively on the lattice either  by projecting on external quark and gluon
states (NPM) as proposed in ref. \cite{NP} or, in the spirit of ref.~\cite{octet}, 
by using chiral Ward Identities \cite{WI,JAPBK}.
Recent studies of the $B$-parameters,
with both non-perturbative renormalization methods, 
 \cite{DS=2}--\cite{contil} and \cite{JAPBK}, show that discretization
effects are less important than those due to the perturbative evaluation
of the mixing coefficients. 

Given the success of the non-perturbative methods in the computation 
of  $B_K$, the NPM has been applied to the 
evaluation of the two $\Delta I=3/2$  $B$-parameters of the electro-penguin 
operators, $B_7^{3/2}$ and $B_8^{3/2}$ (these $B$-parameters coincide 
with those of the operators $O_{4}$ nd $O_{5}$ respectively). Also in this case, as shown
in ref.~\cite{contil}, it has been
found that the non-perturbative renormalization of the lattice operators
gives $B$-parameters that significantly differ from those renormalized
perturbatively~\cite{contil,gupta_bp}.

An extensive study of the renormalization properties of the four-fermion 
operators can be found in \cite{4ferm_teo}. There we detail the 
issues of relevance to the non-perturbative 
renormalization of all the  $\Delta S = 2$ operators. 
We have used these results in the present study.

The NPM for the evaluation of the renormalization constants of lattice
operators consists in imposing suitable renormalization conditions
on lattice amputated quark correlation functions \cite{NP}.
In our case, we compute four-fermion Green functions in the Landau gauge.
All external quark lines are at equal momentum $p$. After amputating
and projecting these correlation functions (see refs.~\cite{DS=2} and
\cite{4ferm_teo} for details), the renormalization conditions are imposed in 
the deep Euclidean region at the scale $p^2 = \mu^2$. This 
renormalization scheme has been recently called the
 Regularization Independent (RI)  scheme~\cite{Ciuchini2} 
(MOM in the early literature) in order to emphasize that the renormalization
conditions are independent of the regularization scheme, although they depend 
on the external states used in the renormalization procedure (and on the
gauge). Thus, at fixed 
cutoff (i.e. fixed $\beta$), we compute non-perturbatively  the renormalization
constants and the renormalized 
operator $\hat O^{RI}(\mu)$ in the RI scheme. In order to obtain the physical
amplitudes, which are renormalization group invariant and scheme 
independent, the renormalized matrix 
elements must subsequently be combined with the corresponding Wilson 
coefficients of the  effective Hamiltonian. For the operators of interest,
the latter are known at the NLO in  continuum perturbation theory \cite{scimemi}. 

In \cite{4ferm_teo}, we have determined non-perturbatively the operator mixing
for the complete basis of four-fermion operators, with the aid of the discrete
symmetries (parity, charge conjugation and switching of flavours),
in the spirit of ref.~\cite{BERNARD2}.
The renormalization of the parity-even operators, relevant to this work,
is more  conveniently expressed in terms of the following basis of five operators:
\bea
Q_1 &=& V \times V + A \times A , \nonumber \\
Q_2 &=& V \times V - A \times A , \nonumber \\
\label{base}
Q_3 &=& S \times S - P \times P , \\
Q_4 &=& S \times S + P \times P , \nonumber \\
Q_5 &=& T \times T .              \nonumber
\eea
 The operators $Q_1,\dots,Q_5$ form a complete basis on the lattice.
In these expressions, $\Gamma \times \Gamma$ (with $\Gamma = V,A,S,P,T$ 
a generic Dirac matrix) stands for $\frac{1}{2}(
\bar \psi_1 \Gamma \psi_2 \bar \psi_3 \Gamma \psi_4 +
\bar \psi_1 \Gamma \psi_4 \bar \psi_3 \Gamma \psi_2) $, where 
$\psi_i,~i=1,\dots,4$ are fermion fields with flavours chosen
so as to reproduce the desired operators (see ref.~\cite{4ferm_teo} for
details). \par
The parity-even parts of the five SUSY operators defined in eq. (\ref{eq:ods2}),
which are the relevant ones for $K^{0}$--$\bar K^{0}$ mixing,   
are related to the operators of eq. (\ref{base}) in the following way:
\bea
O_1 &=&  Q_1 ,\nn \\
O_2 &=&  Q_4 ,\nn \\
O_3 &=&  - \frac{1}{2} (Q_4 - Q_5 ) , \\
O_4 &=&  Q_3 ,\nn \\
O_5 &=&  - \frac{1}{2} Q_2 \ . \nn 
\eea
\par  On the lattice, $Q_1$ mixes under renormalization with 
the other four operators as follows
\be
\label{eq:bk_sub}
\hat Q_1 = Z_{11} \left [ Q_1 + \sum^5_{i=2} Z_{1i} Q_i \right ], \nonumber
\ee
where $Z_{11}$ is a multiplicative logarithmically divergent renormalization
constant; it depends on the coupling and $a\mu$. The mixing coefficients 
$Z_{1i}$ (with $i = 2,\ldots,5$) are finite; they only depend on the lattice
coupling $g_0^2(a)$.

The other renormalized operators are defined as follows:
\begin{eqnarray}
\hat Q_2 &=& Z_{22} Q_2^s + Z_{23} Q_3^s , \nn \\
\hat Q_3 &=& Z_{32} Q_2^s + Z_{33} Q_3^s , \\
\hat Q_4 &=& Z_{44} Q_4^s + Z_{45} Q_5^s , \nn \\
\hat Q_5 &=& Z_{54} Q_4^s + Z_{55} Q_5^s , \nn
\label{eq:q2345ren}
\end{eqnarray}
where the $Z_{ij}$s  are logarithmically divergent
renormalization constants which depend on the coupling and $a\mu$.
The above mixing matrices are not peculiar to the lattice regularization, 
but also occur in the continuum. The breaking of chiral symmetry by the 
Wilson action requires the additional subtractions:
\bea
Q_i^s &=& Q_i + \sum_{j=1,4,5} Z^s_{ij} Q_j, \;\;\; i=2,3 , \nn \\
Q_i^s &=& Q_i + \sum_{j=1,2,3} Z^s_{ij} Q_j, \;\;\; i=4,5 , \nn
\eea
where the $Z^s_{ij}$s are finite coefficients which only depend on $g_0^2(a)$. 

The results for all the renormalization constants $Z_{ij}$ and $Z^s_{ij}$
(computed with the NPM, for  several renormalization scales $\mu$,  at $\beta = 
6.0$ and $6.2$) can be found in \cite{4ferm_teo}.

\section{$B$-parameters}
\label{sec:res}
In this section we discuss the definition of the $B$-parameters and 
their dependence on the renormalization scale. We also sketch the 
extraction of these quantities from  lattice correlation 
functions. \par  The $B$-parameters are usually defined as
\be
B_i(\mu) = 
\frac{ \langle \bar K^0 \vert \hat O_i (\mu) \vert
K^0 \rangle } {\langle \bar K^0 \vert \hat O_i \vert
K^0 \rangle_{VSA} } \,\, ,
\label{eq:bkdef}
\ee
where   the operator matrix elements in the Vacuum Saturation Approximation (VSA)
are given by 
\bea
{\langle \bar K^0 \vert \hat O_1 \vert K^0 \rangle_{VSA} } 
&=& 2 \left ( 1 +  \frac{1}{N_c} \right) 
 \vert \langle \bar  K^0 \vert  \hat  A_\mu \vert 0 \rangle \vert ^2 \,\, , \nn \\
{\langle \bar K^0 \vert \hat O_2 \vert K^0 \rangle_{VSA} } 
&=& -2 \left ( 1 -  \frac{1}{2 N_c} \right) 
 \vert \langle \bar K^0 \vert \hat  P \vert 0 \rangle \vert ^2 \,\, , \nn \\
\label{eq:ds2vsa}
{\langle \bar K^0 \vert \hat O_3 \vert K^0 \rangle_{VSA} } 
&=&
\left ( 1 -  \frac{2}{N_c} \right) 
 \vert \langle \bar K^0 \vert \hat  P \vert 0 \rangle \vert ^2 \,\, , \\
{\langle \bar K^0 \vert \hat O_4 \vert K^0 \rangle_{VSA} } 
&=& 2  \vert \langle \bar K^0 \vert \hat  P \vert 0 \rangle \vert^2
+ \frac{1}{N_c} \vert \langle \bar K^0 \vert \hat A_\mu \vert 0 \rangle \vert^2 
\,\, , \nn \\
{\langle \bar  K^0 \vert \hat O_5 \vert K^0 \rangle_{VSA} } 
&=& \frac{2}{N_c}  \vert \langle \bar  K^0 \vert \hat  P \vert 0 \rangle \vert^2
+  \vert \langle \bar  K^0 \vert \hat A_\mu \vert 0 \rangle \vert^2 
\,\, . \nn 
\eea 
$\hat A_{\mu}$ and $\hat P$ are the renormalized  axial current and pseudoscalar densities, $\hat A_{\mu} = 
Z_{A} A_{\mu} $ and $\hat P = Z_{P} P$,
with $Z_A$ the (finite) renormalization constant of the lattice axial current,
$A_\mu = \bar s \gamma_\mu \gamma_5 d$, and 
$Z_P$ the renormalization constant of the lattice pseudoscalar
density, $P=\bar s \gamma_5 d$. For simplicity, $\hat P$ is renormalized at the same scale 
$\mu$, and in the same renormalization scheme as the four-fermion 
operators (the RI scheme in our case).   Using the relations
\bea  \vert \langle \bar K^0 \vert \hat A_\mu \vert 0 \rangle \vert^2 
&= & M_{K}^{2} f_{K}^{2}  \ , \nonumber \\
 \vert \langle \bar K^0 \vert \hat  P \vert 0 \rangle \vert^2 &=&
 \left( \frac{M_{K}} { m_{s}(\mu) + m_d(\mu) }\right)^{2}
M_{K}^{2} f_{K}^{2}  \ , \label{eq:rels} \eea
where the second equality  is a consequence of the Ward identity for 
the axial current (with  $m_{s}(\mu)$ and $m_{d}(\mu)$  
renormalized in the same scheme and at the same scale  as $\hat P$), we find, with $N_c=3$
\bea \langle  \bar K^{0} \vert  \hat O_{1} (\mu) \vert K^{0} 
\rangle_{VSA} &=&
\frac{8}{3} M_{K}^{2} f_{K}^{2} ,  \nn \\
\langle  \bar K^{0} \vert  \hat O_{2} (\mu) \vert K^{0} \rangle _{VSA} &=&
-\frac{5}{3} \left( \frac{ M_{K} }{ m_{s}(\mu) + m_d(\mu) }\right)^{2}
M_{K}^{2} f_{K}^{2} ,\nn   \\
\langle  \bar K^{0} \vert \hat O_{3} (\mu) \vert K^{0} \rangle _{VSA} &=&
\frac{1}{3} \left( \frac{ M_{K} }{ m_{s}(\mu) + m_d(\mu) }\right)^{2}
M_{K}^{2} f_{K}^{2} ,\label{eq:vsaa} \\
\langle  \bar K^{0} \vert \hat O_{4} (\mu) \vert K^{0} \rangle _{VSA} &=&
2 \left[ \left( \frac{ M_{K} }{ m_{s}(\mu) + m_d(\mu) }\right)^{2}
+ \frac{1}{6} \right]
M_{K}^{2} f_{K}^{2}  ,\nn \\
\langle  \bar K^{0} \vert \hat  O_{5} (\mu) \vert K^{0} \rangle _{VSA} &=&
\frac{2}{3}\left[  \left( \frac{ M_{K} }{ m_{s}(\mu) + m_d(\mu) }\right)^{2}
+ \frac{3}{2} \right] M_{K}^{2} f_{K}^{2}  \nn \  .
 \eea
The VSA values of the matrix elements of $\hat O_{4}$ and $\hat O_{5}$
in eq.~(\ref{eq:vsaa}) differ   from the factors appearing in the 
definition of the $B$-parameters in eq.~(\ref{eq:bpars}) by the terms
proportional to $1/6$ and $3/2$ respectively. These terms, which  originate 
from the squared matrix elements of the axial current in 
eq.~(\ref{eq:ds2vsa}), are of higher order in the chiral 
expansion and  have been dropped in our definition of the $B$-parameters.
This implies that, out of the chiral limit,  the values of 
$B_{4}$ and $B_{5}$  with our definition  differ from those obtained  
by using  eq.~(\ref{eq:vsaa}). To illustrate this point, let us imagine  that   as $m_{s} \to 
\infty$,   for some value of the renormalization scale $\bar \mu$,   
the values of the matrix elements of $\hat O_{4}$ and $\hat O_{5}$  were exactly 
those of the VSA. Under these hypotheses,
using  eq.~(\ref{eq:bpars}), we would get  $B_{4}(\bar 
\mu)=7/6$ and $B_{5}(\bar  \mu)=5/2$ instead of one. 
\par We now explain why we prefer the definition of the $B$-parameters 
given in   eq.~(\ref{eq:bpars}) rather than the standard definition 
obtained from  eq.~(\ref{eq:vsaa}).  Neglecting discretization errors,
the $B$-parameters of the operators $\hat O_{2}$--$\hat O_{5}$ defined in 
eq.~(\ref{eq:bpars}) obey  the renormalization group equation
\be   \mu \frac{ d B_{i}(\mu)}{d\mu} = (\gamma_{O_i} -2 \gamma_{P} ) B_{i}(\mu) \ , \label{eq:rge} \ee
where $\mu d/d\mu = \mu \partial /\partial \mu + \beta(\alpha_{s}) \partial/
\partial \alpha_{s}$, and  $\gamma_{O_i}$ ($\gamma_{O_iO_j}$) and $\gamma_{P}$  are the anomalous dimension
(matrix)  of the operator $\hat O_{i}(\mu)$ and of the scalar density 
respectively~\footnote{ For simplicity we ignore the mixing of the 
operators $\hat O_{2}$--$\hat O_{3}$ and $\hat O_{4}$--$\hat O_{5}$.}.  The physical 
amplitude is given by
\bea \langle  \bar K^{0} \vert {\cal H}_{eff} \vert K^{0} \rangle 
&=& C_{i} (M_{W}/\mu) \langle  \bar K^{0} \vert \hat O_{i}(\mu) \vert K^{0} \rangle 
\nonumber \\ &=& C_{i} (M_{W}/\mu) \times  B_{i}(\mu) \times 
\frac{1}{( m_{s}(\mu) + m_d(\mu)  )^{2}} M_{K}^{4} f_{K}^{2}   \label{eq:schema}\\
&\sim& \Bigl( 
\frac{\alpha_{s}(M_{W})}{\alpha_{s}(\mu)}\Bigr)^{- \gamma_{O_i}/2\beta_{0}}
 \times \Bigl(\alpha_{s}(\mu)\Bigr)^{ (\gamma_{O_i}-2 \gamma_{P})/2 \beta_{0}}
 \times \Bigl(\alpha_{s}(\mu)\Bigr)^{  \gamma_{P}/\beta_{0}}
\nonumber  \ ,  \eea
where, in the last expression,  we have only shown the leading behaviour  of the 
different factors  which depend on $\mu$, namely  the Wilson coefficient, the 
$B$-parameter and the quark masses.  Eq.~(\ref{eq:schema})  shows 
explicitly the cancellation of the $\mu$-dependent terms in the 
amplitude:  the quark masses scale with an anomalous dimension which 
is opposite in sign to that of the pseudoscalar density (since 
$m(\mu) \hat P(\mu)$ is renormalization group invariant) so that 
$B_{i}(\mu) / m^2(\mu)$ scales as the corresponding operator $\hat 
O_{i}(\mu)$; the $\mu$-dependence of the latter is then cancelled by 
that  of the corresponding Wilson coefficient.  This remains true
at all order in $\alpha_s$.
Out of the chiral limit, with  the standard definition of the $B$-parameters  obtained by using 
the VSA matrix elements of  eq.~(\ref{eq:vsaa}), the scaling properties of 
  $B_{4}(\mu)$  and $B_{5}(\mu)$  would have been much more complicated.
 The reason is that, 
in these cases, the two contributions on the right hand side have a 
piece which scales as the squared pseudoscalar density and another 
(proportional to the physical quantity $\vert \langle \bar  K^0 \vert \hat 
A_\mu \vert 0 \rangle \vert^2$) which is renormalization group 
invariant. 
The $\mu$-independence of the final result would then have been 
recovered then in a very intricate way.
Since the definition of the $B$-parameters is conventional, we prefer 
to use that of eq.~(\ref{eq:bpars}),  for which the scaling properties of all the 
$B$-parameters are the simplest ones.  Moreover, with this choice,
 they are the same  as   those  derived in the  chiral limit.
\par In order to extract the $B$-parameters, 
we need to compute the following two- and three-point correlation
functions:
\bea
G_P(t_x,\vec p) = \sum_{\vec x} 
\langle P(x) P^{\dagger} (0) \rangle e^{-\vec p \cdot \vec x} &,&
\,\,\,\,\,\,\,\,\,\,\,
G_A(t_x,\vec p) = \sum_{\vec x} \langle A_0(x) P^{\dagger} (0) \rangle e^{-\vec p 
\cdot \vec x}, 
\label{eq:corrs} \\
G_{\hat O}(t_x,t_y;\vec p, \vec q) &=& \sum_{\vec x,
\vec y} \langle P^{\dagger}(y) \hat O(0) P^{\dagger}(x)\rangle
e^{-\vec p \cdot \vec y} e^{\vec q \cdot \vec x} , \nn
\eea
where $x \equiv (\vec x, t_x), y \equiv (\vec y , t_y)$ and
$\hat O$ stands for any renormalized four-fermion operator of interest. 
All correlation functions have been evaluated with degenerate quark
masses.
By forming suitable ratios of the above correlations, and looking at their
asymptotic behaviour at large time separations, we can isolate the desired
matrix elements
\bea
 R_1 & =& 
                \frac{G_{\hat O_1 }}{Z^2_A G_P G_P} 
             \to \frac{\langle \bar K^0(\vec q)               
           \vert {\hat O_1} \vert K^0(\vec p) \rangle}{
   Z^2_A \vert \langle 0 \vert P \vert K^0 \rangle \vert ^2 }, \nn \\
 R_2 & =& - \frac{1}{2 ( 1 - \frac{1}{2 N_c} ) Z^2_P} 
                  \frac{G_{\hat O_2 }} {G_P G_P} 
            \to - \frac{1}{2 ( 1 - \frac{1}{2 N_c} ) }
                 \frac{\langle \bar K^0(\vec q) 
           \vert {\hat O_2} \vert K^0(\vec p) \rangle}{
   Z^2_P \vert \langle 0 \vert P \vert K^0 \rangle \vert ^2 }, \nn \\
 R_3 & =& \frac{1}{( 1 - \frac{2}{N_c} ) Z^2_P} 
                  \frac{G_{\hat O_3}} {G_P G_P} 
             \to \frac{1}{( 1 - \frac{2}{N_c} ) }
                 \frac{\langle \bar K^0(\vec q) 
           \vert {\hat O_3} \vert K^0(\vec p) \rangle}{
   Z^2_P \vert \langle 0 \vert P \vert K^0 \rangle \vert ^2 }, 
   \label{eq:rapporti} \\
 R_4 & = &  \frac{1}{2 Z^2_P} 
                  \frac{G_{\hat O_4}} {G_P G_P} 
            \to \frac{1}{2}
                 \frac{\langle \bar K^0(\vec q) 
           \vert {\hat O_4} \vert K^0(\vec p) \rangle}{
   Z^2_P \vert \langle 0 \vert P \vert K^0 \rangle \vert ^2 }, \nn \\
 R_5 & =  & \frac{N_c}{2 Z^2_P} 
                  \frac{G_{\hat O_5}} {G_P G_P} 
             \to \frac{N_c}{2}
                 \frac{\langle \bar K^0(\vec q) 
           \vert {\hat O_5} \vert K^0(\vec p) \rangle}{
   Z^2_P \vert \langle 0 \vert P \vert K^0 \rangle \vert ^2 } \ .  \nn 
\eea
We stress that the $B$-parameters extracted from $R_1, R_4$ and $R_5$
are identical to the $B$-parameters  for the operators
$O^{\Delta S = 2}, O^{3/2}_8$ and $O^{3/2}_7$ respectively. 
In ref.~\cite{contil},  
the results referred to the operators  $O^{\Delta S = 2}, O^{3/2}_8$
and $O^{3/2}_7$ at  $\beta=6.0$ only. In this paper, we present the results
for all the $B$-parameters and for $\beta=6.0$ and $6.2$.

\section{Numerical results}
\label{sec:numres}
Our simulations have  been performed at $\beta = 6.0$ and $6.2$ with the tree-level Clover
action, for several values of the quark masses (corresponding  to the 
values of the  hopping parameter $k$  given in table \ref{tab:runs}),
in the quenched approximation. The physical volume is 
approximatively the same on the two lattices. A summary of the main parameters 
 is given in the same table.  
``Time Intervals'' denote the range in time (in lattice units) on which 
the two-point correlation functions have been fitted 
to extract the meson masses and  the matrix elements of $A_{\mu}$ and  
$P$.  The ratios $R_{i}$,  related to  the matrix 
elements of the four-fermion operators,  have been extracted on the  same 
time intervals. Statistical errors have been estimated with the jacknife method,
 by decimating
10 configurations at a time. 
\par As discussed in sec.~\ref{sec:npm}, the renormalization constants have
been obtained from the quark correlation functions, in the Landau gauge.
The results for the $Z$s have been obtained 
on a $16^3 \times 32$  ($16^{3} \times 32$)  lattice at $\beta=6.0$ 
($6.2$), using a statistical sample of  100  (180) configurations.  
In constructing the renormalized operators  we have used the central 
values of the renormalization constants neglecting their 
 statistical errors.  For this reason the errors on the $B$-parameters only 
include those of operator matrix elements.  In the  ratios
(\ref{eq:rapporti}), we also need the  axial-current   renormalization constant $Z_A$ and 
the $\mu$-dependent renormalization constant $Z_P$ of the pseudoscalar density. 
Although $Z_A$ should not depend on $a\mu$, slight variations of its NPM estimate,  arising 
from systematic effects, partially cancel analogous variations of  $R_{1}$, giving  
more stable results  in the extraction of  the matrix elements. 
The NPM estimates for $Z_P$ and $Z_A$ used in the present work are those of
 ref.~\cite{ggrt}.
\begin{table}[h]
\centering
\begin{tabular}{||c|c|c||}
\hline \hline
 & Run A & Run B \\ \hline
$\beta$ & $6.0$ & $6.2$  \\
No. Confs & $460$  & $200$ \\
Volume & $18^{3} \times 64$ & $24^{3} \times 64$ \\ \hline 
& 0.1425 & 0.14144 \\
$k$& 0.1432 & 0.14184 \\
& 0.1440 & 0.14224 \\
& -- & 0.14264  \\ \hline  
Time Intervals & $10$--$22$ & $14$--$26$ \\
& $42$--$54$ & $38$--$50$  \\  \hline 
$a^{-1} (K^{*}) $ (GeV) & $2.12(4)$ & $2.7(1)$ \\
\hline\hline
\end{tabular}
\caption{\it{Summary of the parameters of the runs at $\beta=6.0$ (run 
A) and $\beta=6.2$ (run B). The  calibration of the lattice spacing 
$a^{-1}$ has been done using the lattice-plane method of 
ref.~\protect\cite{giusti}.}} 
\label{tab:runs}
\end{table}

In order to extract the $B$-parameters from the ratios of
eqs.~(\ref{eq:rapporti}), we follow the procedure of ref.~\cite{B_K}, 
by fitting  the $R_{i}$ linearly in $X$ and $Y$ (i.e. linearly in 
$m^{2}_{K}$ and $(p \cdot q)$ ) with the function
\be
\label{eq:r_1}
R_{i} = \alpha_{i} + \beta_{i} X + \gamma_{i}  Y \, ,
\ee
where
\bea
X = \frac{8}{3} \frac{G_A G^\dagger_A}{G_P G_P} \to \frac{8}{3}
\frac{f_K^2 m_K^2}{Z_A^2 \vert \langle 0 \vert P \vert K^0 \rangle \vert ^2}
\, , \quad
Y = \frac{( p \cdot q)}{m^2_K} X  \  , \label{fit_var} \eea with \bea  
p \cdot q  = E(\vec p) E(\vec q)  - \vec p \cdot \vec q  \, ,
\quad \sinh^{2} \left(\frac{E(\vec p)}{2}\right) = \sinh^{2}
\left(\frac{m}{2}\right)  +\sum_{i=1,3} \sin^{2} 
\left(\frac{p_{i}}{2} \right) \ .
\label{fit_var2}
\eea
\par 
Assuming $\alpha_{1}$ and $\beta_{1}$ to be zero, $B_1$
is given by:
\be B_1 = \gamma_{1}\ .  \label{eq:b1} \ee
Since we are working in  the linear approximation in $X$ and $Y$, 
there is no difference  between the value obtained in the 
chiral limit and at the physical kaon mass. 
In the chiral limit, the $B$-parameters for the other four operators are given by
\be
B_{i} = \alpha_i , \qquad ( i = 2,3,4,5) \ .
\label{eq:bc} \ee
We stress again   that $B_4 = B^{3/2}_8$ and $B_5 = B^{3/2}_7$. 
At the physical kaon mass  we have instead
\be
B_{i} = \alpha_i +(\beta_{i} + \gamma_{i}) X_{s} , \qquad ( i = 2,3,4,5) ,
\label{eq:bp} \ee
where $X_{s}$ is obtained by extrapolating linearly $X$ as a function of the 
squared pseudoscalar meson mass to the physical value  $m^{exp}_{K}$.
\par In tables \ref{tab:60} and \ref{tab:62} we give the values of 
the $B$-parameters  in the chiral limit,
extracted using eqs.~(\ref{eq:b1}) and  (\ref{eq:bc}),  
at $\beta=6.0$ and $6.2$ 
respectively; in tables \ref{tab:60p} and \ref{tab:62p} the 
$B$-parameters are evaluated at the physical kaon mass  using 
eq.~(\ref{eq:bp}) for $i=2,3,4,5$.
At $\beta=6.0$, the results for $B_{1}$, $B_{4}$ and $B_{5}$ 
extrapolated to the chiral limit  are slightly  different from those of 
 ref.~\cite{contil}.
There are several  reasons for the differences: 
i) in order to fix the scale and the strange quark mass we have used 
the lattice-plane method of ref.~\cite{giusti}; ii) 
in the present analysis, we use     the ``lattice dispersion 
relation'' of eq.~(\ref{fit_var2}), instead than the continuum one $E^{2}= m^{2} + 
\sum_{i=1,3} p_{i}^{2}$ which was adopted in our previous 
study~\footnote{ Although the continuum and lattice dispersion 
relations are equivalent at the order in $a$ at which we are working,
the latter  gives a better 
description of the data for large momenta, see for example 
ref.~\cite{ukqcdgm}.}; iii) in 
order to reduce the systematic effects due to  higher order terms  in the 
chiral expansion, i.e. to higher powers of  $p \cdot q$, we 
have not used the results corresponding to  $\vec p=2 \pi /L(1,0,0)$  and 
$\vec q=2 \pi/L(-1,0,0)$.  This choice stabilizes the results for $B_{1}$   
between $\beta=6.0$ and $\beta=6.2$  whilst the results for the 
other $B$-parameters remain essentially unchanged. 
\begin{table}
\centering
\begin{tabular}{||c|c|c|c|c|c||}
\hline
\hline
$\mu^2a^2$ & $B_{1}$ & $B_{2}$ & $B_{3}$ & $B_{4}$ & $B_{5}$ \\
\hline
$0.31$ & $0.72 \pm 0.14$ & $0.70 \pm 0.08 $ & $1.62 \pm 0.28$ & $1.21 \pm 0.08$ &
$1.70 \pm 0.24$ \\
$0.62$ & $0.70 \pm 0.15$ & $0.64 \pm 0.04 $ & $1.21 \pm 0.12$ & $1.08 \pm 0.05$ &
$0.81 \pm 0.10$ \\
$0.96$ & $0.70 \pm 0.15$ & $0.61 \pm 0.03 $ & $1.10 \pm 0.08$ & $1.04 \pm 0.04$ &
$0.68 \pm 0.07$ \\
$1.27$ & $0.70 \pm 0.15$ & $0.60 \pm 0.03 $ & $1.03 \pm 0.07$ & $1.01 \pm 0.04$ & 
$0.69 \pm 0.06$ \\
$1.38$ & $0.70 \pm 0.15$ & $0.59 \pm 0.03 $ & $1.03 \pm 0.07$ & $1.01 \pm 0.04$ &
$0.70 \pm 0.05$ \\
$1.85$ & $0.71 \pm 0.15$ & $0.59 \pm 0.03 $ & $0.98 \pm 0.06$ & $0.99 \pm 0.04$ & 
$0.68 \pm 0.05$ \\
$2.46$ & $0.72 \pm 0.15$ & $0.57 \pm 0.03 $ & $0.96 \pm 0.05$ & $1.00 \pm 0.04$ & 
$0.69\pm 0.04$ \\
$4.00$ & $0.74 \pm 0.15$ & $0.55 \pm 0.02 $ & $0.87 \pm 0.04$ & $0.99 \pm 0.03$ & 
$0.73\pm 0.04$ \\
\hline
\hline
\end{tabular}
\caption{\it{Values of the $B$-parameters  in the chiral limit,  at different scales 
$\mu^{2} a^{2}$ for $\beta=6.0$.}}
\label{tab:60}
\end{table}
\begin{table}
\centering
\begin{tabular}{||c|c|c|c|c|c||}
\hline
\hline
$\mu^2a^2$ & $B_{1}$ & $B_{2}$ & $B_{3}$ & $B_{4}$ & $B_{5}$ \\
\hline
$0.31$ & $0.67 \pm 0.21 $ & $0.72 \pm 0.10 $ & $0.90 \pm 0.40$ & $0.99 \pm 0.12$ & 
$0.21 \pm 0.24$ \\
$0.62$ & $0.68 \pm 0.21 $ & $0.63 \pm 0.06 $ & $0.94 \pm 0.16$ & $0.98 \pm 0.08$ & 
$0.46 \pm 0.13$ \\
$0.96$ & $0.68 \pm 0.21 $ & $0.60 \pm 0.04 $ & $0.91 \pm 0.11$ & $0.95 \pm 0.07$ & 
$0.54 \pm 0.10$ \\
$1.27$ & $0.68 \pm 0.21 $ & $0.58 \pm 0.04 $ & $0.89 \pm 0.08$ & $0.93 \pm 0.07$ & 
$0.56 \pm 0.09$ \\
$1.38$ & $0.68 \pm 0.21 $ & $0.58 \pm 0.04 $ & $0.88 \pm 0.08$ & $0.92 \pm 0.07$ & 
$0.55 \pm 0.09$ \\
$1.85$ & $0.67 \pm 0.21 $ & $0.57 \pm 0.03 $ & $0.84 \pm 0.07$ & $0.91 \pm 0.06$ & 
$0.56 \pm 0.08$ \\
$2.46$ & $0.68 \pm 0.21 $ & $0.56 \pm 0.03 $ & $0.82 \pm 0.06$ & $0.91 \pm 0.06$ & 
$0.58 \pm 0.07$ \\
$4.00$ & $0.69 \pm 0.21 $ & $0.54 \pm 0.03 $ & $0.78 \pm 0.05$ & $0.91 \pm 0.06$ & 
$0.60 \pm 0.07$ \\
\hline
\hline
\end{tabular}
\caption{\it{Values of the $B$-parameters  in the chiral limit,  at different scales 
$\mu^{2} a^{2}$ for $\beta=6.2$.}}
\label{tab:62}
\end{table}
\begin{table}
\centering
\begin{tabular}{||c|c|c|c|c|c||}
\hline
\hline
$\mu^2a^2$ & $B_{1}$ & $B_{2}$ & $B_{3}$ & $B_{4}$ & $B_{5}$ \\
\hline
$0.31$ & $0.72 \pm 0.14$ & $0.74 \pm 0.06 $ & $1.57 \pm 0.22$ & $1.19 \pm 0.06$
 & $1.70 \pm 0.19$ \\
$0.62$ & $0.70 \pm 0.15$ & $0.69 \pm 0.03 $ & $1.22 \pm 0.10$ & $1.08 \pm 0.04$
 & $0.92 \pm 0.09$ \\
$0.96$ & $0.70 \pm 0.15$ & $0.66 \pm 0.03 $ & $1.12 \pm 0.07$ & $1.05 \pm 0.03$
 & $0.79 \pm 0.06$ \\
$1.27$ & $0.70 \pm 0.15$ & $0.65 \pm 0.02 $ & $1.06 \pm 0.06$ & $1.03 \pm 0.03$
 & $0.79 \pm 0.05$ \\
$1.38$ & $0.70 \pm 0.15$ & $0.64 \pm 0.02 $ & $1.06 \pm 0.05$ & $1.02 \pm 0.03$
 & $0.79 \pm 0.05$ \\
$1.85$ & $0.71 \pm 0.15$ & $0.63 \pm 0.02 $ & $1.02 \pm 0.05$ & $1.01 \pm 0.03$
 & $0.77 \pm 0.04$ \\
$2.46$ & $0.72 \pm 0.15$ & $0.61 \pm 0.02 $ & $0.99 \pm 0.04$ & $1.02 \pm 0.03$
 & $0.77 \pm 0.04$ \\
$4.00$ & $0.74 \pm 0.15$ & $0.59 \pm 0.02 $ & $0.90 \pm 0.03$ & $1.01 \pm 0.03$
 & $0.81 \pm 0.03$ \\
\hline
\hline
\end{tabular}
\caption{\it{Values of the $B$-parameters  at the physical kaon mass,  at different scales 
$\mu^{2} a^{2}$ for $\beta=6.0$.}}
\label{tab:60p}
\end{table}
\begin{table}
\centering
\begin{tabular}{||c|c|c|c|c|c||}
\hline
\hline
$\mu^2a^2$ & $B_{1}$ & $B_{2}$ & $B_{3}$ & $B_{4}$ & $B_{5}$ \\
\hline
$0.31$ & $0.67 \pm 0.21 $ & $0.75 \pm 0.07 $ & $1.05 \pm 0.29$ & $1.05 \pm 0.09$
 & $0.57 \pm 0.18$ \\
$0.62$ & $0.68 \pm 0.21 $ & $0.66 \pm 0.04 $ & $0.98 \pm 0.12$ & $1.01 \pm 0.06$
 & $0.67 \pm 0.10$ \\
$0.96$ & $0.68 \pm 0.21 $ & $0.63 \pm 0.03 $ & $0.95 \pm 0.08$ & $0.99 \pm 0.06$
 & $0.70 \pm 0.08$ \\
$1.27$ & $0.68 \pm 0.21 $ & $0.61 \pm 0.03 $ & $0.92 \pm 0.07$ & $0.97 \pm 0.05$
 & $0.71 \pm 0.07$ \\
$1.38$ & $0.68 \pm 0.21 $ & $0.61 \pm 0.03 $ & $0.91 \pm 0.06$ & $0.97 \pm 0.05$
 & $0.70 \pm 0.07$ \\
$1.85$ & $0.67 \pm 0.21 $ & $0.60 \pm 0.03 $ & $0.88 \pm 0.06$ & $0.97 \pm 0.05$
 & $0.70 \pm 0.06$ \\
$2.46$ & $0.68 \pm 0.21 $ & $0.59 \pm 0.03 $ & $0.86 \pm 0.05$ & $0.97 \pm 0.05$
 & $0.71 \pm 0.06$ \\
$4.00$ & $0.69 \pm 0.21 $ & $0.57 \pm 0.02 $ & $0.82 \pm 0.04$ & $0.97 \pm 0.05$
 & $0.73 \pm 0.06$ \\
\hline
\hline
\end{tabular}
\caption{\it{Values of the $B$-parameters  at the physical kaon mass,  at different scales 
$\mu^{2} a^{2}$ for $\beta=6.2$.}}
\label{tab:62p}
\end{table}
\par We now describe the criteria followed in order to obtain our best estimates of the $B$-parameters.
Although  we have data at two different values of the lattice spacing, the 
statistical errors, and the uncertainties in the extraction of the 
matrix elements,  are too large to enable any extrapolation to the continuum limit 
$a \to 0$ : within the precision of our results  we cannot detect the dependence of
$B$-parameters on $a$. For this reason,  we estimate the central 
values by  averaging the $B$-parameters obtained with the physical mass
$m_K^{exp}$ at the two values of 
$\beta$. Since the results at $\beta=6.0$ have  smaller statistical 
errors but suffer from larger discretization  effects, we do not 
weight the averages with the quoted statistical 
errors but take simply the sum of the two values divided by two. As 
far as the errors are concerned we take the largest of the two statistical 
errors. This is a rather conservative way of estimating the 
errors.  In order to compare the results of Run A and Run B, we must choose the same 
physical renormalization scale  $\mu$. 
Using  the estimates of the lattice spacing given in table 
\ref{tab:runs},  we have taken $\mu^{2} a^{2}=0.96$ and $\mu^{2} 
a^{2}=0.62$, corresponding to $\mu=2.08$ GeV and $\mu=2.12$ GeV, at 
$\beta=6.0$ and $6.2$ respectively.  We quote the results as obtained 
at $\mu=2$ GeV, since the running of the matrix elements between $\mu 
\sim 2.1$ and $2.0$ is totally negligible in comparison with the 
final errors. In table \ref{tab:summary}, we summarize the values 
which have been used to give the final estimates. The columns denoted
by $m_K^{exp}$ have been used to get the final results in
eq.~(\ref{eq:fres}). 
 \begin{table}[hptb]
\centering
\begin{tabular}{||c|c|c|c|c|c||}\hline\hline
$B(\mu \simeq 2~GeV)$ 
& $m_{K}=0$   
& $m_{K}=0$   
& $m_{K}=m_{K}^{exp}$ 
&  $m_{K}=0$  
& $m_{K}=m_{K}^{exp}$     \\
&$\beta=6.0$&$\beta=6.0$&$\beta=6.0$&$\beta=6.2$  &$\beta=6.2$\\
& ref.\protect\cite{contil}   &this work& this work&this work&this work
   \\
\hline \hline
$B_1$   & 0.66(11) & 0.70(15) & 0.70(15)& 0.68(21)& 0.68(21) \\
\hline
$B_2$   & ---      &  0.61(3) & 0.66(3) & 0.63(6) & 0.66(4) \\
\hline
$B_3$   & ---      & 1.10(8)  & 1.12(7) & 0.94(16) & 0.98(12) \\
\hline
$B_4$   & 1.03(3)  &  1.04(4) & 1.05(3) & 0.98 (8) & 1.01(6)\\
\hline
$B_5$   & 0.72(5)  & 0.68(7)  & 0.79(6) & 0.46(13) & 0.67(10) \\
\hline \hline
\end{tabular}
\caption{\it{$B$-parameters at the renormalization scale 
$\mu = a^{-1} \simeq 2$~GeV, corresponding to $\mu^{2} a^{2}=0.96$ and $\mu^{2} 
a^{2}=0.62$  at  $\beta=6.0$ and $6.2$ respectively. All results are in 
the RI (MOM) scheme.}}
\label{tab:summary}
\end{table}
\par
In ref.~\cite{gupta_bp} $B_2$ and $B_3$ have been obtained at $\beta = 6.0$
with the Wilson action and the operators renormalized perturbatively in the
$\MSbar$ scheme; the result is
\bea
B_2 = 0.59 \pm 0.01 \nonumber \\
B_3 = 0.79 \pm 0.01
\label{eq:Bpt}
\eea
Although a direct comparison is not possible (our results are in the RI
scheme), to the extent that the matching coefficients between the two schemes
are a small effect \cite{contil}, comparison of eqs.(\ref{eq:fres}) and
(\ref{eq:Bpt}) suggests that perturbative renormalization gives significantly
different results in some cases.
This confirms the need for non perturbative renormalization.

\section{Conclusions}
\label{sec:concl}

In this paper we have presented a lattice calculation of the matrix elements
of the most  general set of $\Delta S=2$ dimension-six  four-fermion
operators, renormalized non-perturbatively in the RI (MOM) 
scheme~\cite{NP}--\cite{4ferm_teo}.  The calculations have been 
performed at two different values of the lattice spacing $a$. Although 
our precision is not sufficient to make an 
extrapolation to the continuum limit, the comparison between the 
results on two different lattices  allows a better estimate of the 
final errors.
The main  results for the five $B$-parameters are summarized  in 
tab.~\ref{tab:summary}.  From this table, we have extracted  our best 
estimates which are given in eq.~(\ref{eq:fres}).
We observe that the  lattice  values  of  $B_{3,4}$ are close to their VSA 
whereas  this is not true for $B_{1,2,5}$.  
Our  results  allow an improvement in  the accuracy  of   phenomenological analyses 
intended to put bounds on basic parameters of theories
beyond the Standard Model.

\section*{Acknowledgements}
We thank I. Scimemi for discussions during the completion of this work
and T.~Bhattacharya R.~Gupta and S.R.~Sharpe for pointing us out their
results of ref.~\cite{gupta_bp}.
L.C., G.M. and A.V. acknowledge partial support by the M.U.R.S.T.; 
V.G. acknowledges the partial support by CICYT under grant number  AEN-96/1718.  
M.T. acknowledges the PPARC for support through grant no.  GR/L22744
and the INFN for partial support. A.~D. acknowledges the INFN for partial support. 
C.R.A. acknowledges support from the Nuffield Foundation.

\end{document}